\documentclass[pra,showpacs,showkeys,twocolumn,floatfix]{revtex4}

\usepackage{amsmath}
\usepackage{amsfonts}            
\usepackage{amssymb}

\usepackage{mathrsfs}

\usepackage{eufrak}

\usepackage{graphicx}
\usepackage{epstopdf}

\usepackage{dcolumn}
\usepackage{bm}

\begin{document}

\title{Fine-tuning molecular energy levels by nonresonant laser pulses}

\author{Mikhail Lemeshko}

\author{Bretislav Friedrich}

\affiliation{%
Fritz-Haber-Institut der Max-Planck-Gesellschaft, Faradayweg 4-6, D-14195 Berlin, Germany
}%

\date{\today}

\begin{abstract}

We evaluate the shifts imparted to vibrational and rotational levels of a linear molecule by a nonresonant laser field at intensities of up to $10^{12}$ W/cm$^2$. Both types of shift are found to be either positive or negative, depending on the initial rotational state acted upon by the field. An adiabatic field-molecule interaction imparts a rotational energy shift which is negative and exceeds the concomitant positive vibrational shift  by a few orders of magnitude. The rovibrational states are thus pushed downward in such a field. A nonresonant pulsed laser field that interacts nonadiabatically with the molecule is found to impart rotational and vibrational shifts of the same order of magnitude. The nonadiabatic energy transfer occurs most readily at a pulse duration which amounts to about a tenth of the molecule's rotational period, and vanishes when the sudden regime is attained for shorter pulses. We applied our treatment to the much studied $^{87}$Rb$_2$ molecule in the last bound vibrational levels of its lowest singlet and triplet electronic states.  Our calculations indicate that $15$~ns and $1.5$~ns laser pulses of an intensity in excess of $5\times10^9$~W/cm$^2$ are capable of dissociating the molecule due to the vibrational shift. Lesser shifts can be used to fine tune the rovibrational levels and thereby to affect collisional resonances by the nonresonant light. The energy shifts due to laser intensities of $10^9$~W/cm$^2$ may be discernible spectroscopically, with a 10 MHz resolution.

\end{abstract}

\maketitle

\section{Introduction}
\label{intro}

The hybridization of the rotational states of an anisotropic molecule by a nonresonant laser field exerts chiefly a twofold effect on the molecule's energy levels: (i) depending on the initial rotational state, the field can impart angular momentum to the molecule or remove it, which alters the centrifugal term in the molecule's electronic potential and hence pushes its vibrational and rotational manifolds upward or downward; (ii) the hybridization can both increase and decrease the energy of the rotational state with respect to the host vibrational level and thus reinforce or contravene the effect of the upward or downward push by the effective potential. The twofold effect may play a significant role in fine-tuning atomic collisional resonances since, in that case, a reference energy level is provided by the atomic states whose response to the nonresonant radiative field differs both qualitatively and quantitatively from that of the molecular states. Prompted by the current work on photo- and magneto-asscociation of ultracold atoms~\cite{KreStwFrieColdMolecules} we undertook to examine and map out the twofold effect systematically.

In our previous work related to weakly bound (diatomic) molecules, we showed that a nonresonant laser field can be used to probe near-threshold vibrational states by ``shaking''~\cite{LemFriPRL09}: the anisotropic polarizability interaction with a nonresonant laser field imparts a critical value of angular momentum fine-tuned to push a rotationless vibrational level over the threshold for dissociation. The imparted angular momentum converts into the molecule's libration -- hence ``shaking.'' In a complementary study we found an accurate analytic approximation that allows to readily evaluate the critical angular momentum needed~\cite{LemFriPRArapid09, LemFriJAMS10}. Here we deal with the case of rotational levels supported by a vibrational state which remains bound by the effective potential.

We resort to the problem's time-dependent Schr\"{o}dinger 
equation to sample the key observables, such as energy, angular momentum, and alignment. We do so in different temporal and interaction-strength regimes, which range from the adiabatic to the sudden limit, and from the weak-field to the strong-field limit, respectively. While the angular momentum and alignment have been tapped before both in the adiabatic~\cite{FriHerPRL95, FriHerJPC95} and non-adiabatic~\cite{OrtigosoFriedrich99,CaiFriedrichPRL01,CaiFriedrichCCCC01,StapelfeldtSeidemanRMP03} regimes, the energy has not. Yet, it is the expectation value of the Hamiltonian that is key to the effect sub (ii) and hence, in combination with the effect sub (i), to the ability to fine-tune collisional resonances by nonresonant laser pulses. We treat the problem in reduced, dimensionless variables, which allows us to sample the behavior of any (diatomic) molecule in a nonresonant radiative field. Finally, we evaluate the effect feasible nonresonant laser fields, both cw and pulsed, are expected to have on the levels of the much examined $^{87}$Rb$_2$ molecule in its $X^1\Sigma$ and $a^3\Sigma$ electronic states.

\section{Interaction of an anisotropic molecule with nonresonant laser pulses}

As in the previous treatments of the interaction of a linear molecule with nonresonant laser pulses~\cite{OrtigosoFriedrich99,CaiFriedrichPRL01,CaiFriedrichCCCC01}, we limit our considerations to a pulsed plane-wave radiation of frequency $\nu$ and time profile $g(t)$ and assume the oscillation frequency to be far removed from any molecular resonance and much higher than either the inverse pulse duration, $\tau^{-1}$, or the rotational period, $\tau_{r}$. Averaged over the rapid oscillations, the effective Hamiltonian, $H(t)$, lacks any permanent dipole term  and its time dependence is reduced to that of the time profile:
\begin{equation}
\label{ham1}
H(t)= B \left[ \mathbf{J}^{2} - g(t) \left(\Delta \omega \cos ^{2}\theta +\omega _{\bot
} \right)  \right]
\end{equation}
The interaction is characterized by the dimensionless anisotropy parameter $\Delta \omega \equiv \omega _{||}-\omega _{\bot }$, where $\omega _{||,\bot } \equiv 2\pi \alpha_{||,\bot }I/(Bc)$, with $\alpha _{||}(r)$ and $\alpha _{\bot }(r)$ the polarizability components parallel and perpendicular to the molecular axis, $B$ the rotational constant, and $I$ the laser intensity. Because of the azimuthal symmetry about the field vector, the induced dipole potential involves just the polar angle $\theta$ between the molecular axis and the polarization plane of the laser pulse. For homonuclear molecules, the dependence of the polarizability anisotropy, $\Delta \alpha (r)\equiv \alpha _{||}(r)-\alpha _{\bot }(r)$, on the internuclear distance, $r$, is well captured at large $r$ by Silberstein's expansion, $\Delta \alpha (r)=6\alpha^2_0 r^{-3}+6\alpha_0^3 r^{-6}+...$, with $\alpha_0$ the atomic polarizability~\cite{Silberstein17}. We note that higher-order terms in the multipole expansion of the potential, such as second-order hyperpolarizability, pertaining to the 4th power of the field strength, are likely negligible at laser intensities below $10^{12}$ W/cm$^2$, see, e.g., ref.~\cite{RenardPRL03}.

By diving through $B/\pi$, the time-dependent Schr\"{o}dinger equation pertaining to Hamiltonian (\ref{ham1}) becomes dimensionless,
\begin{equation}
\label{schr1}
i \frac{\pi \hbar }{B}\frac{\partial \psi (t)}{\partial t}=\pi \frac{H(t)}{B}\psi (t)
\end{equation}
and clocks the time in terms of the rotational period $\tau_r = \pi \hbar/B$. The energy is expressed in terms of the rotational constant $B$. The solutions of eq.~(\ref{schr1}) can be expanded in field-free rotor wavefunctions $|JM\rangle $ (pertaining to eigenenergies 
$E_{J}$),
\begin{equation}
\psi (\Delta \omega (t))=\sum_{J}c_{J}(\Delta \omega (t))|J,M\rangle \exp
\left( -\frac{iE_{J}t}{\hbar }\right)
\end{equation}
whose time-dependent coefficients, $c_{J}\left( \Delta \omega \left(
t\right) \right) \equiv c_J(t)$, solely determine the solutions at given initial
conditions (in the interaction representation). The \emph{hybridization} coefficients $c_{J}$ can be
found from the set of differential equations 
\begin{multline}
i\frac{\hbar }{B}\stackrel{.}{c}_{J}(t)=-\sum_{J^{\prime }}c_{J^{\prime
}}(t)\left\langle J,M\right| \Delta \omega \cos ^{2}\theta +\omega _{\bot
}\left| J^{\prime },M\right\rangle \\
\times  \exp \left[ -\frac{i\left( E_{J^{\prime
}}-E_{J}\right) t}{\hbar }\right] g(t)
\label{diff1}
\end{multline}
which reduces to a tridiagonal form on account of the non-vanishing matrix elements $\left\langle
J,M\right| \cos ^{2}\theta \left| J^{\prime },M\right\rangle $ and can be solved by standard methods~\cite{OrtigosoFriedrich99,CaiFriedrichCCCC01}. Note that in a linearly polarized laser field, $M$ remains a good quantum number.

\begin{figure}
\includegraphics[width=8cm]{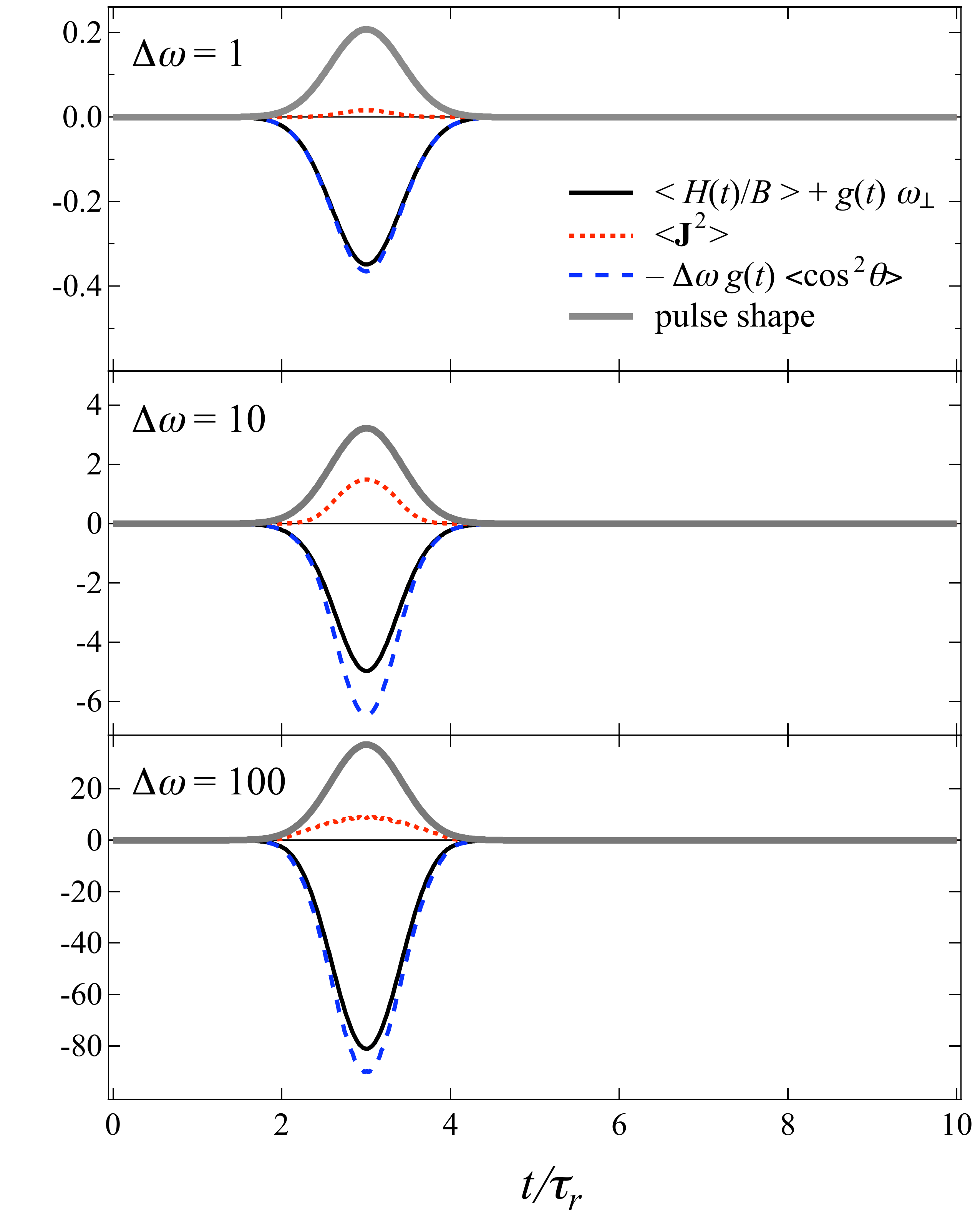}
\caption{\label{fig:E_J2_cos2_tau1_J00} Expectation values of the Hamiltonian, $\langle H(t)/B \rangle + g(t)\omega_\perp$ (black solid curve), of the angular momentum, $\langle \mathbf{J}^2 \rangle$ (red dotted curve), and of the alignment cosine term, $- \Delta \omega g(t) \langle \cos^2\theta \rangle$ (blue dashed curve), for the $\vert 0, 0\rangle$ state subjected to a pulse of duration $\tau = 1~\tau_r$.  The rotational period, $\tau_r = \pi \hbar/B$, is used as a unit of time. The pulse shape (in arbitrary units) is shown by thick grey curve. See text.}
\end{figure}

\begin{figure}
\includegraphics[width=8cm]{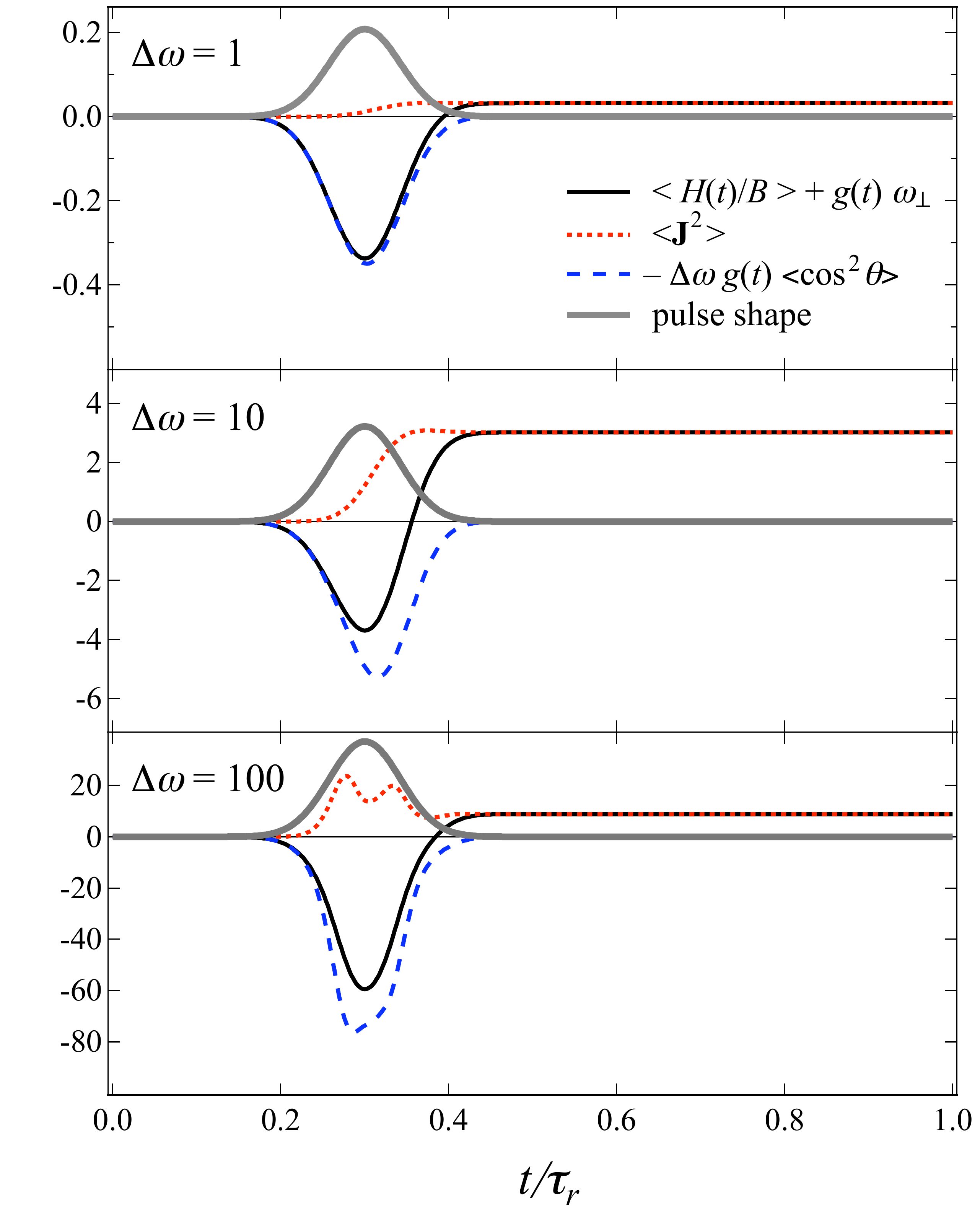}
\caption{\label{fig:E_J2_cos2_tau01_J00} Expectation values of the Hamiltonian, $\langle H(t)/B \rangle + g(t)\omega_\perp$ (black solid curve), of the angular momentum, $\langle \mathbf{J}^2 \rangle$ (red dotted curve), and of the alignment cosine term, $- \Delta \omega g(t) \langle \cos^2\theta \rangle$ (blue dashed curve), for the $\vert 0, 0\rangle$ state subjected to a pulse of duration $\tau = 0.1~\tau_r$. The rotational period, $\tau_r = \pi \hbar/B$, is used as a unit of time. The pulse shape (in arbitrary units) is shown by thick grey curve. See text.}
\end{figure}

We consider the pulse shape function to be a Gaussian, $g(t)=\exp \left[-4\ln(2)t^{2}/\tau^{2} \right]$, characterized by a full width at half maximum, $\tau$, the ``pulse duration.''  If the laser pulse duration is longer than the rotational period, $\tau \gtrsim \tau_r$, the interaction is adiabatic and the molecule behaves as if the field were static at any instant. The states thereby created are the stationary pendular states~\cite{FriHerPRL95, FriHerJPC95}. For $\tau \lesssim \tau_r$, the time evolution is nonadiabatic and the molecule ends up in a rotational wavepacket~\cite{OrtigosoFriedrich99}. The wavepacket comprises a finite number of free-rotor states and thus may recur after the pulse has passed, giving rise to hybridization under field-free conditions. We label the molecular states by the good quantum number $M$ and the nominal value $\tilde{J}$, which corresponds to the rotational quantum number $J$ of the field free rotor state the molecule occupied before the laser pulse struck.

\begin{figure}
\includegraphics[width=8cm]{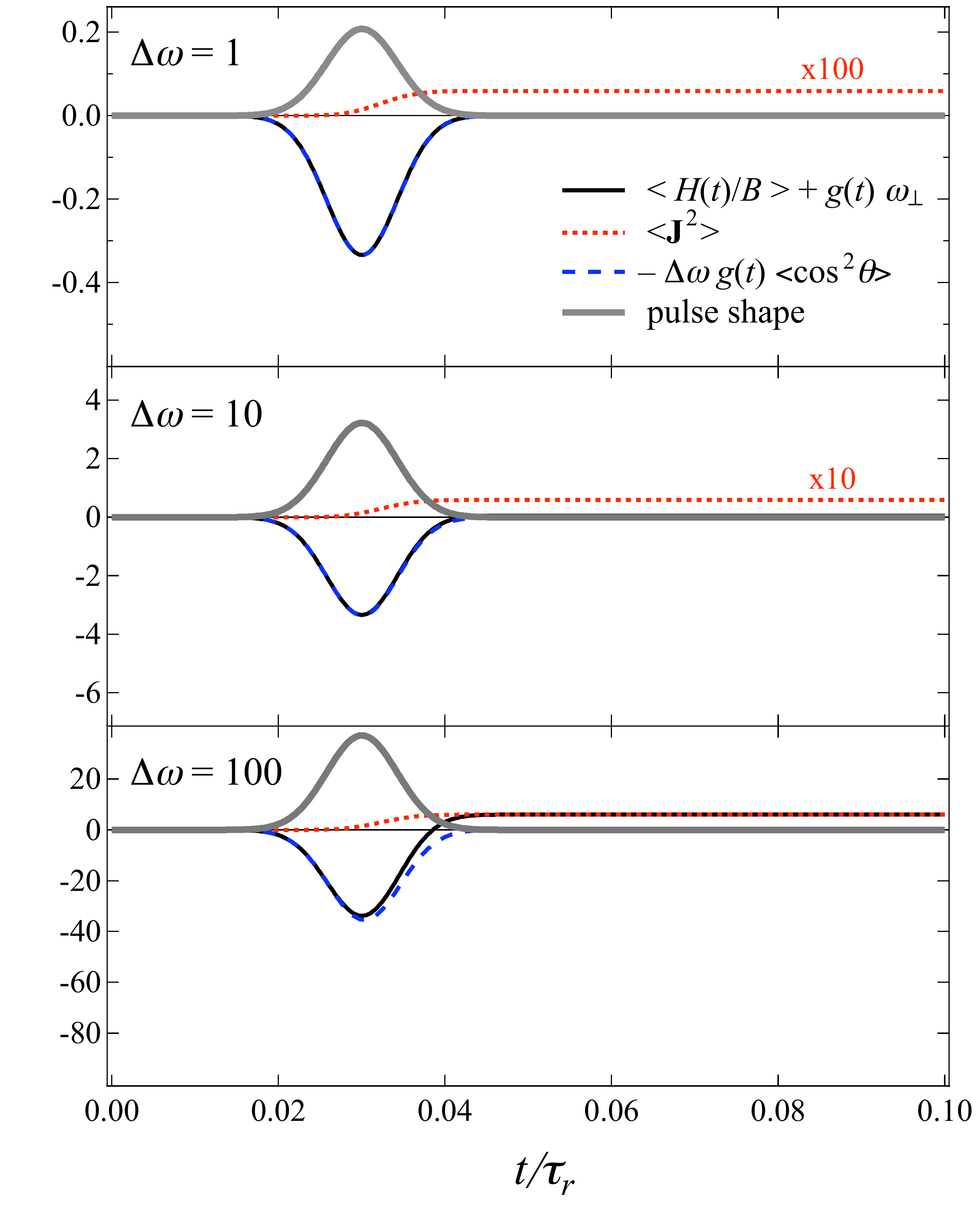}
\caption{\label{fig:E_J2_cos2_tau001_J00} Expectation values of the Hamiltonian, $\langle H(t)/B \rangle + g(t)\omega_\perp$ (black solid curve), of the angular momentum, $\langle \mathbf{J}^2 \rangle$ (red dotted curve), and of the alignment cosine term, $- \Delta \omega g(t) \langle \cos^2\theta \rangle$ (blue dashed curve), for the $\vert 0, 0\rangle$ state subjected to a pulse of duration $\tau = 0.01~\tau_r$. The rotational period, $\tau_r = \pi \hbar/B$, is used as a unit of time. The pulse shape (in arbitrary units) is shown by thick grey curve. See text.}
\end{figure}

\begin{figure}
\includegraphics[width=8cm]{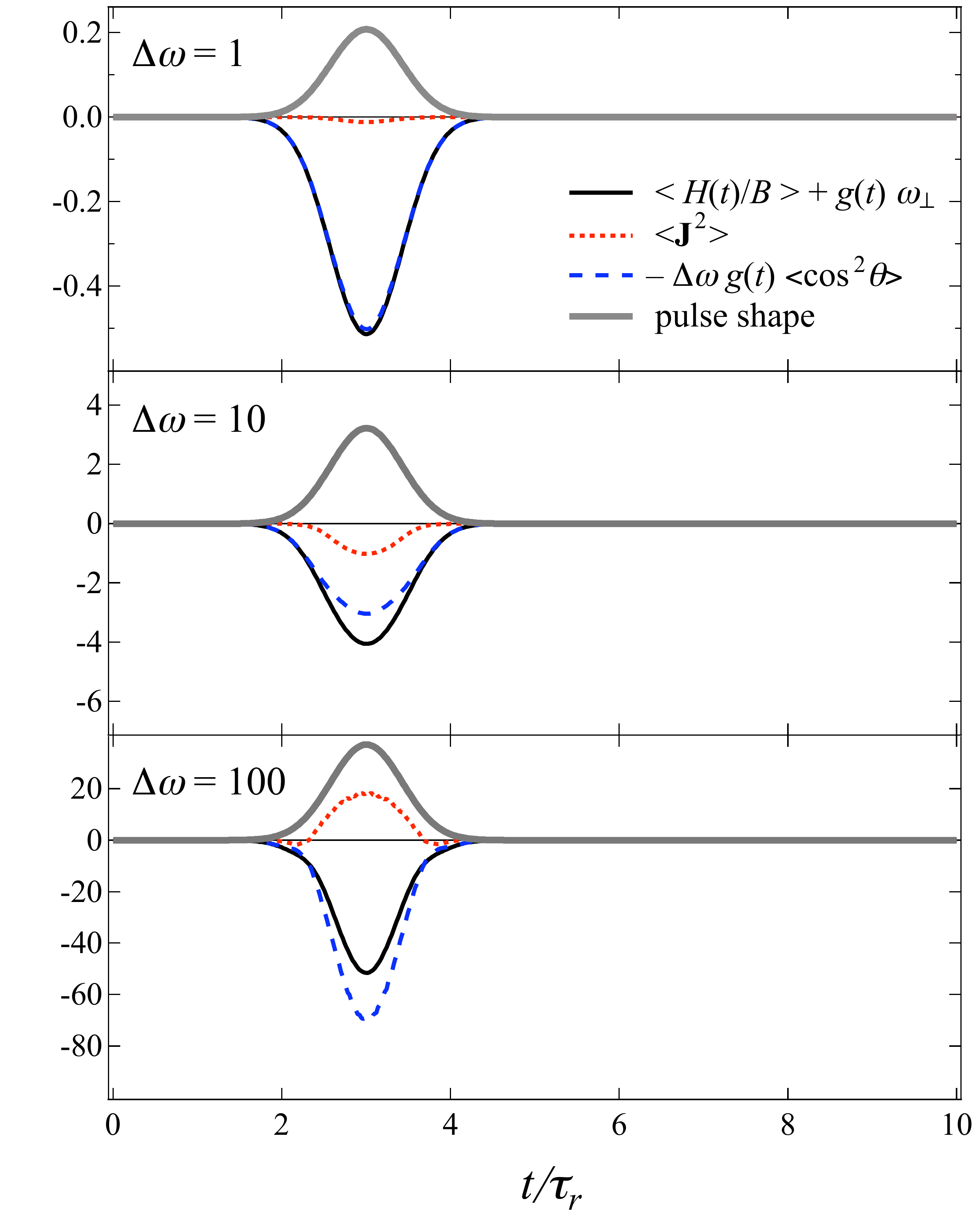}
\caption{\label{fig:E_J2_cos2_tau1_J20} Expectation values of the Hamiltonian, $\langle H(t)/B \rangle + g(t)\omega_\perp$ (black solid curve), of the angular momentum, $\langle \mathbf{J}^2 \rangle$ (red dotted curve), and of the alignment cosine term, $- \Delta \omega g(t) \langle \cos^2\theta \rangle$ (blue dashed curve), for the $\vert 2, 0\rangle$ state subjected to a pulse of duration $\tau = 1~\tau_r$. The rotational period, $\tau_r = \pi \hbar/B$, is used as a unit of time. The pulse shape (in arbitrary units) is shown by thick grey curve. The field-free energy and  the $\langle \mathbf{J}^2 \rangle$ values are subtracted from the quantities shown. See text.}
\end{figure}

\begin{figure}
\includegraphics[width=8cm]{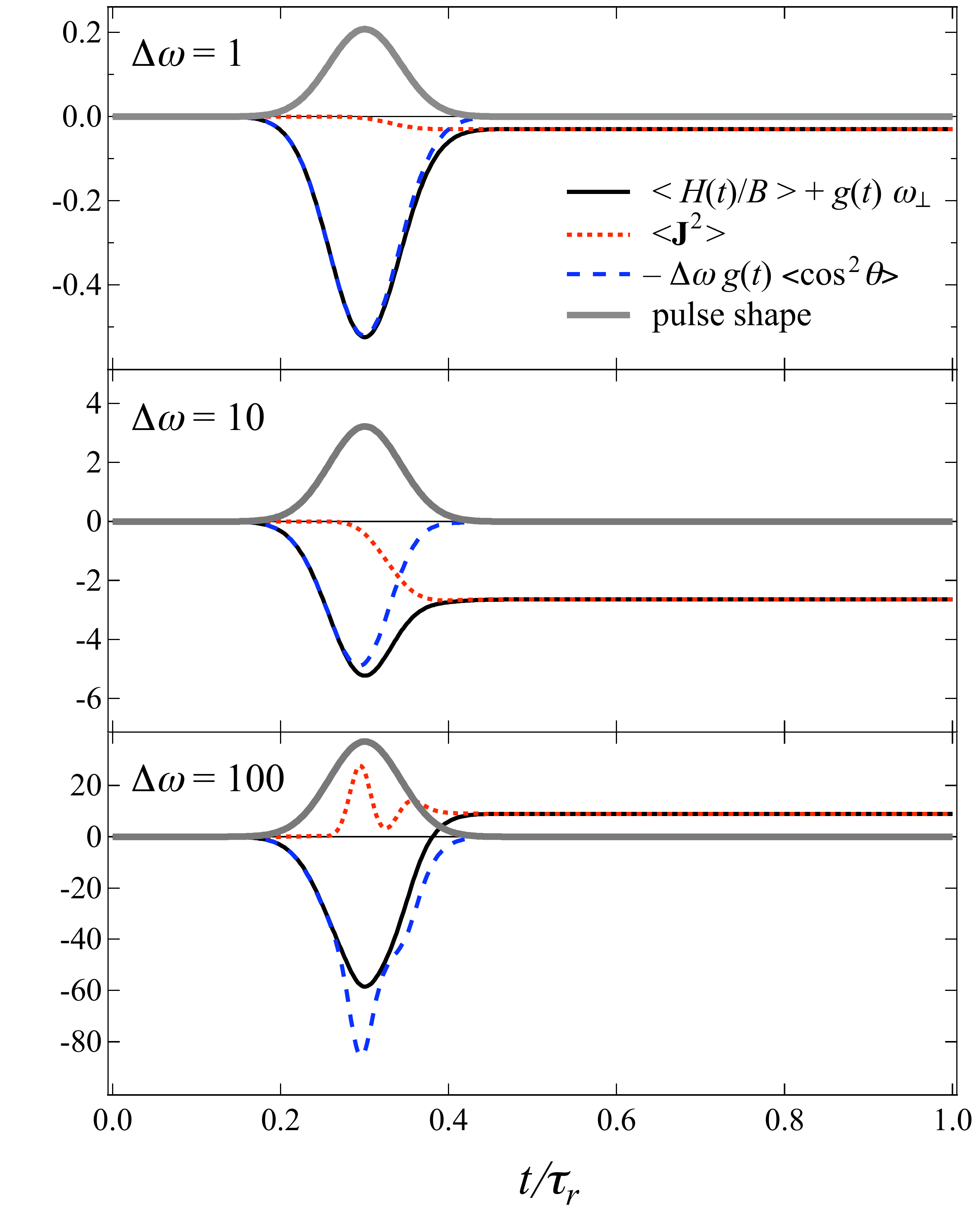}
\caption{\label{fig:E_J2_cos2_tau01_J20} Expectation values of the Hamiltonian, $\langle H(t)/B \rangle + g(t)\omega_\perp$ (black solid curve), of the angular momentum, $\langle \mathbf{J}^2 \rangle$ (red dotted curve), and of the alignment cosine term, $- \Delta \omega g(t) \langle \cos^2\theta \rangle$ (blue dashed curve), for the $\vert 2, 0\rangle$ state subjected to a pulse of duration $\tau = 0.1~\tau_r$.  The rotational period, $\tau_r = \pi \hbar/B$, is used as a unit of time. The pulse shape (in arbitrary units) is shown by thick grey curve. The field-free energy and  the $\langle \mathbf{J}^2 \rangle$ values are subtracted from the quantities shown. See text.}
\end{figure}

\begin{figure}
\includegraphics[width=8cm]{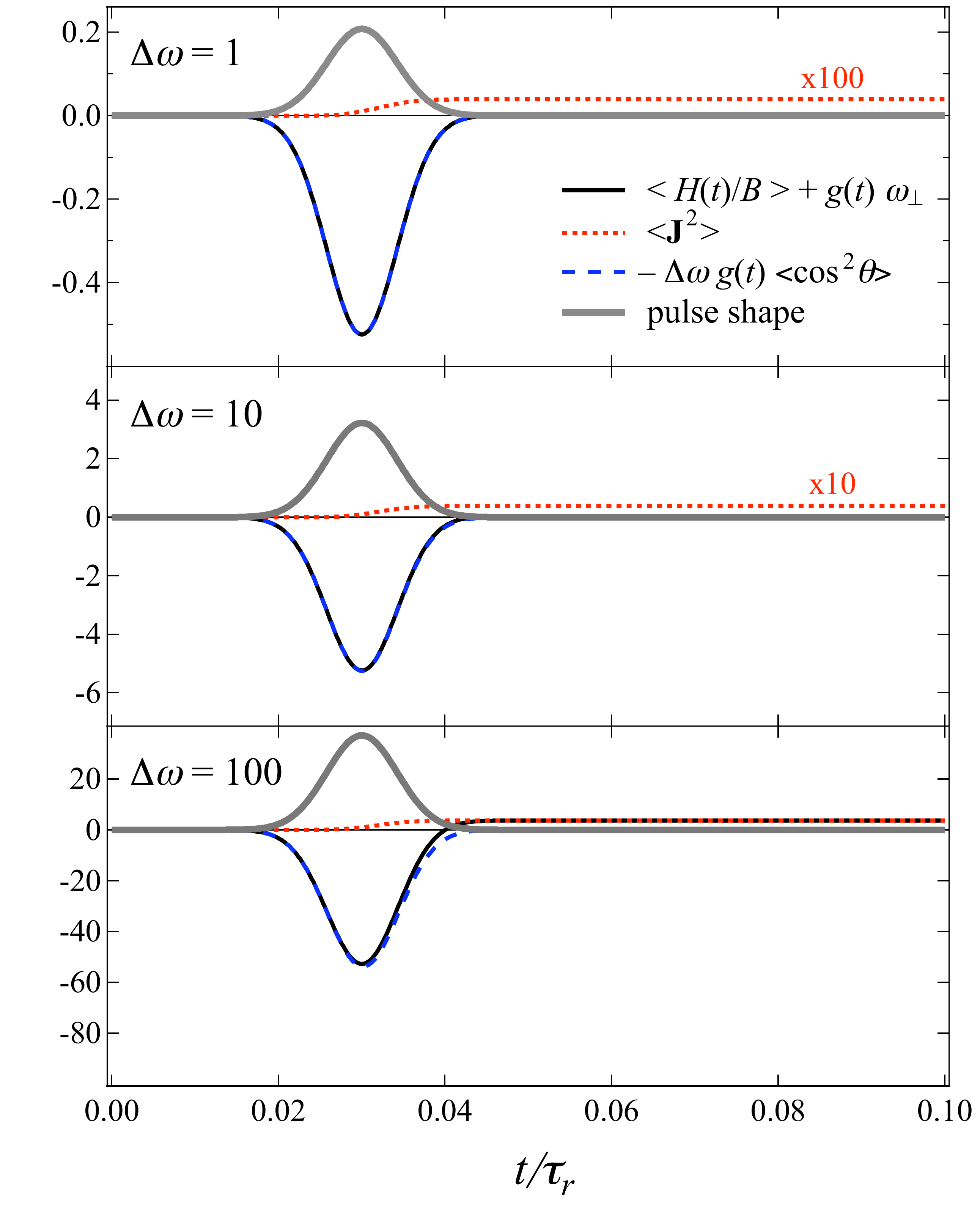}
\caption{\label{fig:E_J2_cos2_tau001_J20} Expectation values of the Hamiltonian, $\langle H(t)/B \rangle + g(t)\omega_\perp$ (black solid curve), of the angular momentum, $\langle \mathbf{J}^2 \rangle$ (red dotted curve), and of the alignment cosine term, $ - \Delta \omega g(t) \langle \cos^2\theta \rangle$ (blue dashed curve), for the $\vert 2, 0\rangle$ state subjected to a pulse of duration $\tau = 0.01~\tau_r$. The rotational period, $\tau_r = \pi \hbar/B$, is used as a unit of time. The pulse shape (in arbitrary units) is shown by thick grey curve. The field-free energy and  the $\langle \mathbf{J}^2 \rangle$ values are subtracted from the quantities shown. See text.}
\end{figure}

\begin{figure}
\includegraphics[width=8cm]{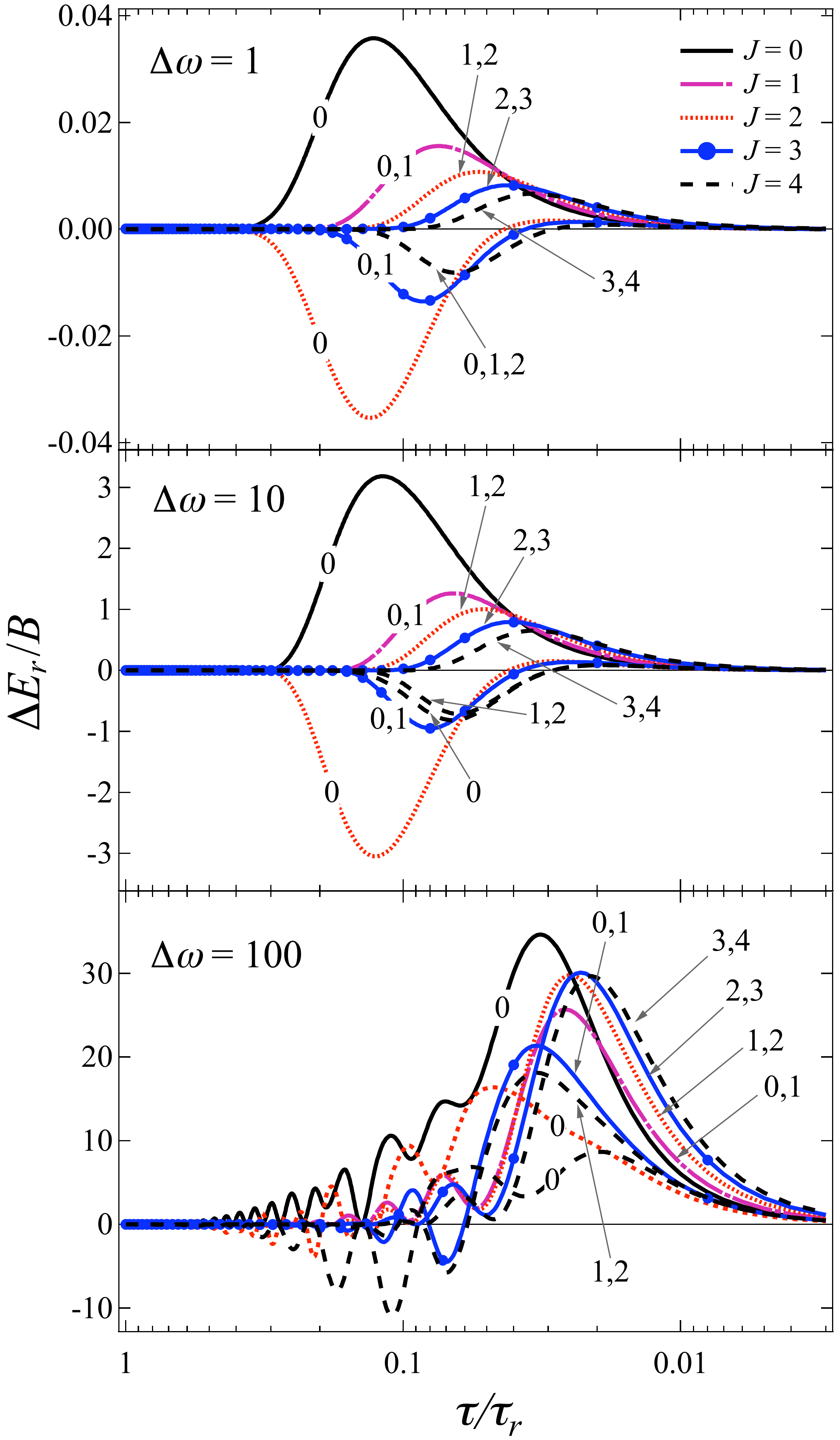}
\caption{\label{fig:DeltaE} Rotational energy transferred to a molecule by a nonresonant laser pulse as a function of the pulse duration $\tau$, expressed in units of the rotational period $\tau_r = \pi \hbar/B$. Different colors and curve styles correspond to different values of $\tilde{J}$; the numbers  show the values of $|M|$.}
\end{figure}

\begin{figure}
\includegraphics[width=8cm]{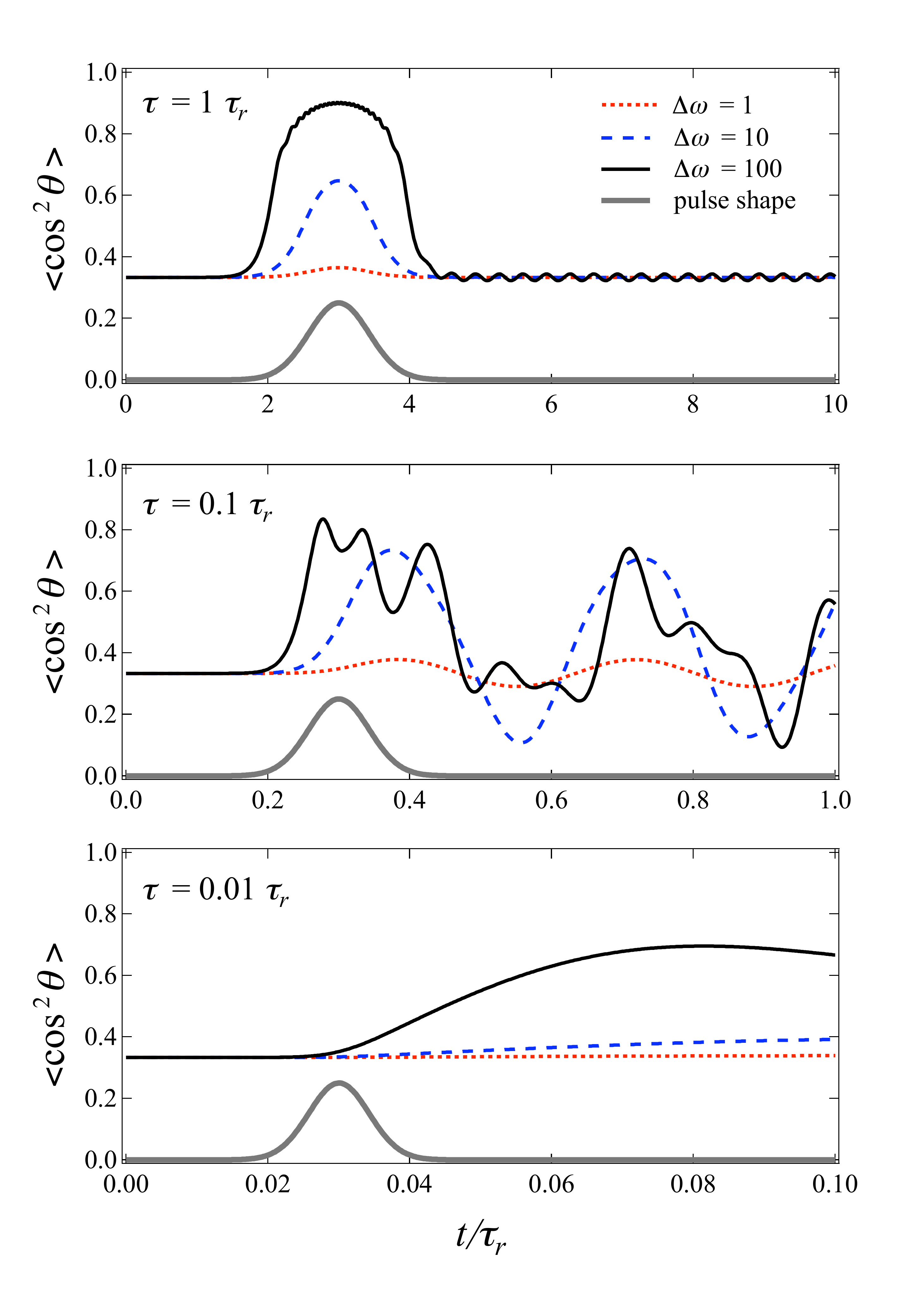}
\caption{\label{fig:cos2} Expectation values of the alignment cosine $\langle \cos^2\theta \rangle$ as a function of time expressed in units of the rotational period $\tau_r = \pi \hbar/B$ for different pulse durations $\tau$ and field intensities $\Delta \omega$.}
\end{figure}

\begin{figure}[t]
\includegraphics[width=8cm]{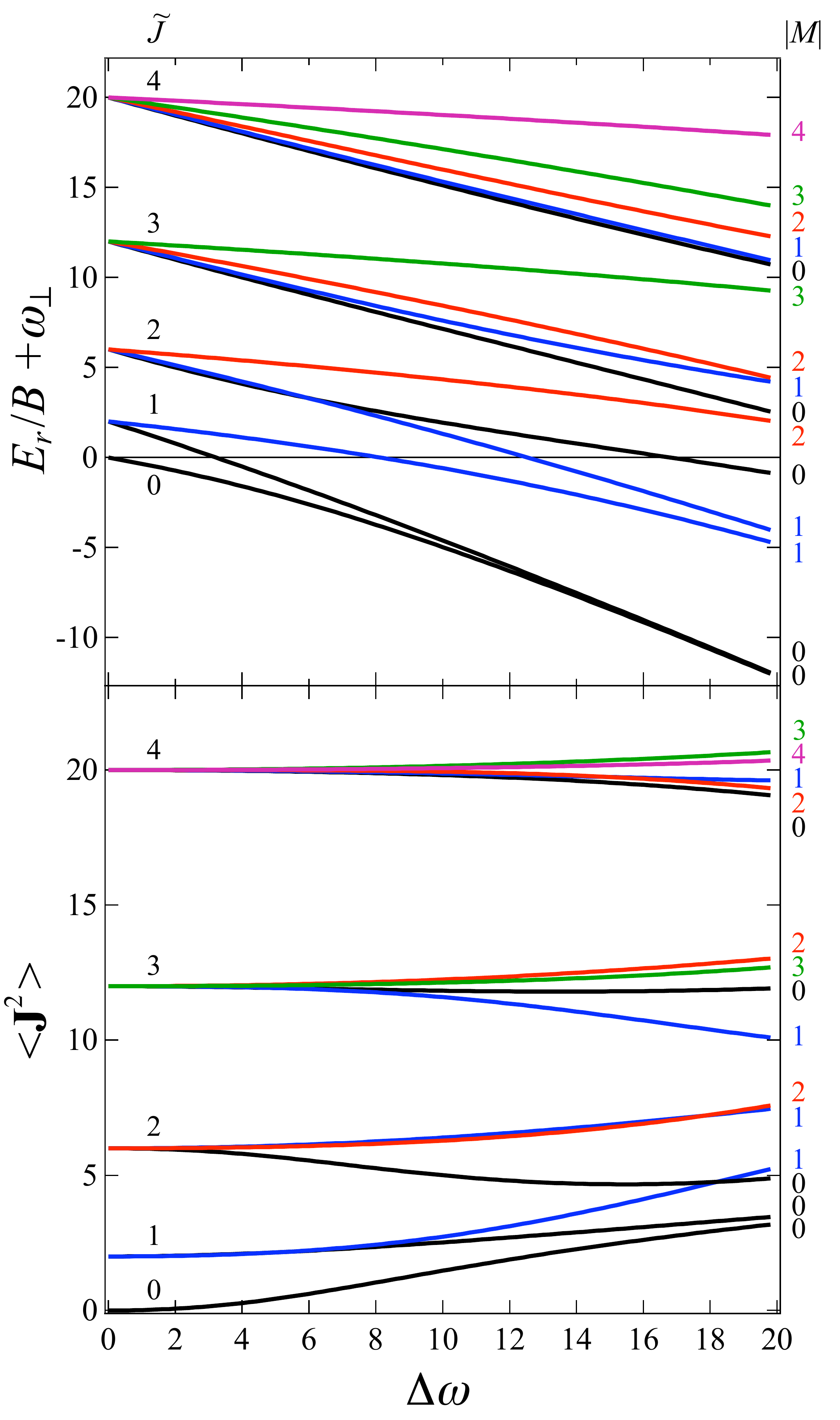}
\caption{\label{fig:cw_laser_effect} Eignenergies (upper panel) and expectation values of squared angular momentum (lower panel) for states with $\tilde{J}=0 - 4$ of a linear molecule in a cw laser field. Thin black line shows the zero of energy. The eigenenergies exhibit an additional field-dependent shift of $-\omega_\perp$, which is not included in the upper panel. }
\end{figure}

The reduced time-dependent expectation value of Hamiltonian (\ref{ham1}), which represents the reduced time-dependent average energy of the rotational wavepacket,
\begin{equation}
	\label{HamilExpValueGeneral}
	\frac{\langle H(t) \rangle}{B}  = \langle \mathbf{J}^2 (t)\rangle - \Delta \omega g(t) \langle \cos^2 \theta (t)\rangle - \omega_\perp g(t),
\end{equation}
is comprised of terms dependent on the expectation values of the squared angular momentum, $\langle \mathbf{J}^2 (t)\rangle$, and of the alignment cosine, $\langle \cos^2 \theta (t)\rangle$. Since the value of $\omega_\perp$ is contingent upon a particular molecule, it is more convenient to work with $\langle H(t) \rangle/B+ \omega_\perp g(t)$ rather than the expectation value $\langle H(t) \rangle/B$ itself. The time-dependent expectation values of the imparted angular momentum and of the alignment cosine are given, respectively, by
\begin{equation}
	\label{J2ExpValue}
	\langle \mathbf{J}^2(t) \rangle_{\tilde{J} M}  = \sum_{J'} c^2_{J'}(t) J'(J'+1)- \tilde{J}(\tilde{J}+1)
\end{equation}
and
\begin{equation}
	\label{AlignCos}
	\langle \cos^2 \theta (t) \rangle_{\tilde{J} M}  = \sum_{J',J''} c_{J'}(t) c_{J''}^*(t) \langle J'', M \vert \cos^2 \theta \vert J', M \rangle
\end{equation}
where the matrix elements of the $\cos^2 \theta$ operator are given by
\begin{multline}
	\label{Cos2matrel}
	\langle J'', M \vert \cos^2 \theta \vert J', M \rangle = \frac{1}{3}\delta_{J',J''} \\
	+ \frac{2}{3} \sqrt{\frac{2 J'+1}{2J''+1}} C(J' 2 J'', M 0 M) C(J' 2 J'', 0 0 0),
\end{multline}
with $C(J_1 J_2 J, M_1 M_2 M)$ the Clebsch-Gordan coefficients and $\delta_{J',J''}$ the Kronecker delta. Note that only matrix elements with $J''-J' = 0; \pm 2$ are non-vanishing.

The noninteger value of the angular momentum $\langle \mathbf{J}^2 \rangle$ imparted by the laser field contributes to the effective electronic potential,
\begin{equation}
	\label{EffectivePot}
	W(r) = V(r)+ \frac{J(J+1) \hbar^2}{2 m r^2}+ \frac{\langle \mathbf{J}^2 \rangle \hbar^2}{2 m r^2}=U(r) + \frac{\langle \mathbf{J}^2 \rangle \hbar^2}{2 m r^2},
\end{equation}
where $V(r)$ is the Born-Oppenheimer molecular potential with $m$ the molecule's reduced mass and $U(r)=V(r)+ J(J+1) \hbar^2/(2 m r^2)$ is the field-free effective potential. The imparted centrifugal term,  $\langle \mathbf{J}^2 \rangle \hbar^2/(2 m r^2)$, thus alters the position of the field free vibrational levels and shifts the entire rotational manifold they host. The energy shift, $\Delta E_v$, due to the laser field of a vibrational level $v$ is given by the difference of its binding energy in the potentials $W(r)$ and $U(r)$,
\begin{equation}
\label{Ev}
\Delta E_v = E_v [W(r)] - E_v [U(r)].
\end{equation}
The binding energies $E_v [W(r)]$ and $E_v [U(r)]$ are obtained by solving the radial Schr\"{o}dinger equation for $W(r)$ and $U(r)$, cf. ref.~\cite{LemFriPRL09}. This is the effect sub (i) mentioned in Section \ref{intro}.

The effect sub (ii) is given by the imparted rotational energy
\begin{equation}
\label{Er}
\Delta E_r = \langle H(t \rightarrow \infty) \rangle  - B  J(J+1).
\end{equation}

The overall energy shift, $\Delta E_t$, of a field-free rotational state  $|J,M\rangle $ is then 
\label{Et}
\begin{equation}
\Delta E_t = \Delta E_v + \Delta E_r
\end{equation}
It can be both an upward or downward shift as either $\Delta E_v$ or $\Delta E_r$ can be positive or negative. The following section examines these shifts.

\section{Behavior of time-dependent expectation values}

Figures~\ref{fig:E_J2_cos2_tau1_J00}--\ref{fig:E_J2_cos2_tau001_J00} show the time evolution of  the expectation values of the $\langle \mathbf{J}^2(t) \rangle$ and $-\Delta \omega g(t) \langle \cos^2 \theta (t)\rangle$ terms that make up Hamiltonian~(\ref{HamilExpValueGeneral}) as well as of the Hamiltonian $\langle H(t) \rangle/B+ \omega_\perp g(t)$  itself for $\vert 0,0 \rangle$ chosen as the initial state. Fig.~\ref{fig:E_J2_cos2_tau1_J00} pertains to the adiabatic regime, characterized by $\tau \ge \tau_r$. One can see that both $\langle \mathbf{J}^2 (t)\rangle$ and $- \Delta \omega g(t) \langle \cos^2 \theta (t)\rangle$ simply follow the laser-pulse shape, leading to a lowering of energy during the pulse. This is also known as the AC Stark effect. Fig.~\ref{fig:E_J2_cos2_tau01_J00} pertains to the nonadiabatic regime, characterized by $\tau < \tau_r$. It shows that the molecule keeps a fraction of the imparted angular momentum $\langle \mathbf{J}^2 (t)\rangle$ even after the pulse has passed. The resulting wavepacket thus ends up with a nonzero average energy, which can only be removed by a subsequent perturbation. The average energy consists solely of the angular momentum term, $\langle \mathbf{J}^2 (t)\rangle$, as the alignment contribution (not the alignment itself, see below), $- \Delta \omega g(t) \langle \cos^2 \theta (t)\rangle$, vanishes, due to the $g(t)$ term, after the pulse has passed.  Fig.~\ref{fig:E_J2_cos2_tau001_J00} pertains to a highly nonadiabatic case, characterized by $\tau \ll \tau_r$. The behavior of the wavepacket qualitatively resembles that for the  $\tau = 0.1~\tau_r$ case, but the fraction of the energy imparted to the molecule is much less. It is worth noting that in the sudden limit, $\langle \mathbf{J}^2 (t)\rangle$ vanishes altogether, as one can show analytically by making use of the wavepacket given by eq. (13) of ref. \cite{CaiFriedrichCCCC01}. 

Figures~\ref{fig:E_J2_cos2_tau1_J20}--\ref{fig:E_J2_cos2_tau001_J20} show the time evolution of $\langle \mathbf{J}^2(t) \rangle$, $-\Delta \omega g(t) \langle \cos^2 \theta (t)\rangle$, and $\langle H(t) \rangle/B+ \omega_\perp g(t)$ in different temporal regimes for the $\vert 2,0 \rangle$ initial state. The effect of the field on the $\vert 2,0 \rangle$ state is seen to be somewhat different from the previously considered case of the $\vert 0,0 \rangle$ state. The difference arises chiefly because the rotational wavepacket now comprises not only states with higher angular momenta, $J>2$, but also a state with a lower angular momentum, namely $J=0$. As a result, for weak field strengths, $\Delta \omega = 1$ and $10$, the hybridization with the close-lying $J=0$ state predominates and the expectation value of the angular momentum decreases. In the nonadiabatic regime, $\tau = 0.1~\tau_r$ and $\tau = 0.01~\tau_r$, the molecule's energy and $\langle \mathbf{J}^2 \rangle$ is lower after the pulse has passed than before its arrival. For stronger fields, $\Delta \omega = 100$, the hybridization with higher $J$'s becomes dominant and positive values of angular momentum and of energy are imparted to the molecule.

\begin{table*}
\centering
\caption{Parameters of the last bound vibrational states of the $^{87}$Rb$_2$ molecule in its lowest singlet and triplet electronic states: the binding energy $E_b$, rotational constant $B$, rotational period $\tau_r$, polarizability anisotropy $\Delta \alpha$, polarizability perpendicular component $\alpha_\perp$, and the critical angular momentum needed to dissociate the molecule, $\langle \mathbf{J}^{*2} \rangle$.}
\vspace{0.2cm}
\label{table:Param}
\begin{tabular}{| c | c | c | c | c |  c | c |}
\hline 
State & $E_b$~($10^{-4}$cm$^{-1}$) & $B$~($10^{-4}$cm$^{-1}$) & $\tau_r$~(ns) & $\Delta \alpha$~(\AA$^3$) & $\alpha_\perp$~(\AA$^3$) & $\langle \mathbf{J}^{*2} \rangle$  \\[5pt]
\hline
$X^1\Sigma~v=124, \vert 0, M \rangle$  & $10.16$  & $1.16$ & 144 & 0.132 & 48.85 & 5.16 \\[5pt]
$X^1\Sigma~v=124, \vert 1, M \rangle$  & $6.42$  & $1.10$ & 152 & 0.122 & 48.85  &  3.16 \\[12pt]
$a^3\Sigma~v=40, \vert 0, M \rangle$  &  $8.35$ & $1.09$ & 153 & 0.128 & 48.85 & 4.55 \\[5pt]
$a^3\Sigma~v=40, \vert 1, M \rangle$  &  $4.39$ & $0.99$ & 168 & 0.117 & 48.85 & 2.55  \\[5pt]
\hline
\end{tabular}
\end{table*}

\begin{table*}
\centering
\caption{Effect of a nonresonant cw laser field of intensity $I = 5 \times 10^8$~W/cm$^2$ on the last bound vibrational states of the $^{87}$Rb$_2$ molecule in its lowest singlet and triplet electronic states. The table lists the dimensionless interaction parameters $\Delta \omega$ and  $\omega_\perp$, alignment cosine $\langle \cos^2\theta \rangle$, imparted angular momentum $\langle \mathbf{J}^{2} \rangle$, rotational shift $\Delta E_r$, vibrational shift $\Delta E_v$, and the total energy shift $\Delta E_t$. Rotational states are labeled by $\vert \tilde{J}, |M| \rangle$. Long laser pulses, $\tau \gtrsim \tau_r$, lead to the same behavior.}
\vspace{0.2cm}
\label{table:cw}
\begin{tabular}{| c | c | c | c | c | c | c | c |}
\hline 
State & $\Delta \omega$ &  $\omega_\perp$ & $\langle \cos^2\theta \rangle$ & $\langle \mathbf{J}^{2} \rangle$ & $\Delta E_r$~($10^{-4}$cm$^{-1}$) & $\Delta E_v$~($10^{-4}$cm$^{-1}$) & $\Delta E_t$~($10^{-4}$cm$^{-1}$) \\[5pt]
\hline
$X^1\Sigma~v=124, \vert 0, 0 \rangle$  & 5.67  & 2100.34 & 0.526 & 0.56 & $-2445$  & $1.16$ & $-2444$ \\[5pt]
$X^1\Sigma~v=124, \vert 1, 0 \rangle$  & 5.53 & 2224.8 & 0.669 & 0.18 & $-2444$ & $0.93$  & $-2443$ \\[5pt]
$X^1\Sigma~v=124, \vert 1, 1 \rangle$  & 5.53  & 2224.8  & 0.263 & 0.19 & $-2442$ & $0.93$ & $-2441$ \\[15pt]
$a^3\Sigma~v=40, \vert 0, 0 \rangle$  & 5.83 & 2230.87 & 0.531  & 0.59 & $-2445$ & $1.20$ & $-2444$  \\[5pt]
$a^3\Sigma~v=40, \vert 1, 0 \rangle$  & 5.88  & 2456.98  & 0.673 & 0.21 & $-2444$ & $0.41$ & $-2444$ \\[5pt]
$a^3\Sigma~v=40, \vert 1, 1 \rangle$  & 5.88 & 2456.98 & 0.268 & 0.21 & $-2442$ & $0.41$ & $-2442$ \\[5pt]
\hline
\end{tabular}
\end{table*}

\begin{table*}
\centering
\caption{Effect of a pulsed nonresonant laser field of intensity $I = 5 \times 10^9$~W/cm$^2$ on the last bound vibrational states of the $^{87}$Rb$_2$ molecule in its lowest singlet and triplet electronic states. The pulse duration is $\tau = 15\text{ ns} \approx 0.1~\tau_r$. The table lists the dimensionless interaction parameter $\Delta \omega$, maximum alignment cosine $\langle \cos^2\theta \rangle_\text{max}$,  imparted angular momentum $\langle \mathbf{J}^{2} \rangle$, rotational shift $\Delta E_r$,  vibrational shift  $\Delta E_v$, and the total energy shift $\Delta E_t$. Rotational states are labeled by $\vert \tilde{J}, |M| \rangle$.}
\vspace{0.2cm}
\label{table:Short}
\begin{tabular}{| c | c | c | c | c | c | c |}
\hline 
State & $\Delta \omega$ &   $\langle \cos^2\theta \rangle_\text{max}$ & $\langle \mathbf{J}^{2} \rangle$ & $\Delta E_r$~($10^{-4}$cm$^{-1}$) & $\Delta E_v$~($10^{-4}$cm$^{-1}$) & $\Delta E_t$~($10^{-4}$cm$^{-1}$) \\[5pt]
\hline
$X^1\Sigma~v=124, \vert 0, 0 \rangle$  & 56.7   & 0.705 & 7.40 &  $8.6$  & $> E_b$ & $> E_b$  \\[5pt]
$X^1\Sigma~v=124, \vert 1, 0 \rangle$  &  55.3 & 0.835  & 2.49  &  $2.7$ &  $5.3$ & $8.0$  \\[5pt]
$X^1\Sigma~v=124, \vert 1, 1 \rangle$  &  55.3 & 0.504 & 2.49 & $2.7$  & $5.3$   & $8.0$  \\[15pt]
$a^3\Sigma~v=40, \vert 0, 0 \rangle$  &  58.3  &  0.219  & 7.73 & $8.2$   &  $>E_b$  & $>E_b$  \\[5pt]
$a^3\Sigma~v=40, \vert 1, 0 \rangle$  &  58.8  & 0.808  & 2.29 & $2.3$ & $3.9$ & $6.2$ \\[5pt]
$a^3\Sigma~v=40, \vert 1, 1 \rangle$  &  58.8 & 0.456 & 2.29  & $2.3$ &  $3.9$ & $6.2$  \\[5pt]
\hline
\end{tabular}
\end{table*}

\begin{table*}
\centering
\caption{Effect of a pulsed nonresonant laser field of intensity $I = 5 \times 10^9$~W/cm$^2$ on the last bound vibrational states of the $^{87}$Rb$_2$ molecule in its lowest singlet and triplet electronic states. The pulse duration is $\tau = 1.5\text{ ns} \approx 0.01~\tau_r$. The table lists values of the dimensionless interaction parameter $\Delta \omega$, maximum alignment cosine $\langle \cos^2\theta \rangle_\text{max}$, angular momentum $\langle \mathbf{J}^{2} \rangle$  imparted to the molecule, rotational shift $\Delta E_r$,  vibrational shift $\Delta E_v$, and the total energy shift $\Delta E_t$. Rotational states are labeled by $\vert \tilde{J}, |M| \rangle$.}
\vspace{0.2cm}
\label{table:VeryShort}
\begin{tabular}{| c | c | c | c | c | c | c |}
\hline 
State & $\Delta \omega$ &   $\langle \cos^2\theta \rangle_\text{max}$ & $\langle \mathbf{J}^{2} \rangle$ & $\Delta E_r$~($10^{-4}$cm$^{-1}$) & $\Delta E_v$~($10^{-4}$cm$^{-1}$) & $\Delta E_t$~($10^{-4}$cm$^{-1}$) \\[5pt]
\hline
$X^1\Sigma~v=124, \vert 0, 0 \rangle$  & 56.7   & 0.636 & 2.11  & $2.5$   &  $4.6$ &  $7.1$ \\[5pt]
$X^1\Sigma~v=124, \vert 1, 0 \rangle$  & 55.3  & 0.800 & 2.22  & $2.4$  & $4.9$  & $7.3$ \\[5pt]
$X^1\Sigma~v=124, \vert 1, 1 \rangle$  & 55.3  & 0.434 & 2.22  & $2.4$ & $4.9$ & $7.3$ \\[15pt]
$a^3\Sigma~v=40, \vert 0, 0 \rangle$  &  58.3  & 0.628   & 1.98 & $2.2$ & $3.9$  & $6.1$ \\[5pt]
$a^3\Sigma~v=40, \vert 1, 0 \rangle$  &  58.8  & 0.795 & 2.06 & $2.0$ & $4.4$ & $6.4$  \\[5pt]
$a^3\Sigma~v=40, \vert 1, 1 \rangle$  & 58.8  & 0.388  & 2.06 & $2.0$ & $4.4$ & $6.4$  \\[5pt]
\hline
\end{tabular}
\end{table*}

Figure~\ref{fig:DeltaE} shows how the imparted rotational energy, $\Delta E_r$, depends on the pulse duration $\tau$ and the interaction strength $\Delta \omega $. Increasing from zero in the adiabatic regime, $\tau \sim \tau_r$, the imparted energy passes through a maximum or minimum at a pulse duration of $\tau \sim 0.1~\tau_r$, and subsequently approaches zero again when the process becomes highly nonadiabatic, $\tau \sim 0.01~\tau_r$. For moderate field strengths, $\Delta \omega \le 10$, the largest positive $\Delta E_r$ occurs for the $\vert 0,0\rangle$ state, while the negative one occurs for the  $\vert 2,0\rangle$ state. At high interaction strengths, $\Delta \omega = 100$, the hybridization with higher-lying $J$ states becomes dominant and the imparted energy is chiefly positive for all states. 

The fast oscillations of $\Delta E_r$ at $\Delta \omega=100$, discernible also in the temporal dependence of $\langle \mathbf{J}^{2}\rangle$ at $\Delta \omega=100$ shown in Figs. \ref{fig:E_J2_cos2_tau01_J00} and \ref{fig:E_J2_cos2_tau01_J20}, arise in the intermediate temporal regime between the adiabatic and sudden limits as $\tau/\tau_r$ changes from about $5$ to about $0.05$. In either limit, the wavefunction is a smooth, non-oscillatory function of time when the field is on: trivially so in the adiabatic limit, but equally so in the sudden limit, where the wavefunction is shaped by a ``once and for all'' impulsive transfer of action from the field to the molecule and exhibits a $\tau \Delta \omega$ scaling, cf. \cite{CaiFriedrichCCCC01}. In the intermediate regime, however, the wavefunction is all but monotonous and its oscillations reflect the molecule's attempts to minimize its energy in the course of a rotational period.
  
Figure~\ref{fig:cos2} shows the time evolution of the alignment cosine in different temporal and interaction strength regimes, as defined by the values of $\tau$ and $\Delta \omega$. One can see that in the adiabatic regime, $\tau \sim \tau_r$, the alignment cosine simply follows the pulse shape, as a result of which a molecule in the $\vert 0,0 \rangle$ state is not aligned after the laser pulse has passed. This is not the case in the nonadiabatic regime, $\tau = 0.1~\tau_r$ and $\tau = 0.01~\tau_r$, where the molecule exhibits  ``revivals,'' i.e., periodically recurring alignment after the waning of the laser pulse.

We note that for laser pulses shorter than the vibrational period, the imparted energy, angular momentum, and alignment will depend on the internuclear distance $r$ at which the molecule is struck by the laser pulse. In order to take this into account, the above observables have to be scaled according to $\Delta \omega = 2\pi \Delta \alpha(r) I/[B(r) c]$, where the polarizability anisotropy $\Delta \alpha=\Delta \alpha(r)$ and rotational constant $B=B(r)$ are no longer expectation values but $r$-dependent quantities, cf. ref.~\cite{LemFriPRL09}. For large values of $r$, the Silberstein expansion yields the dependence $\Delta \omega \propto I/r$.

On the other hand, in the adiabatic limit, $\tau \gg \tau_r$, the time profile becomes constant, $g(t)=1$, corresponding to a cw laser field. The Schr\"odinger equation~(\ref{schr1}) reduces to an eigenvalue problem, whose solutions are the pendular states~\cite{FriHerPRL95, FriHerJPC95},
\begin{equation}
\psi (\Delta \omega (t))=\sum_{J}c_{J}(\Delta \omega (t))|J,M\rangle \equiv \vert \tilde{J}, M, \Delta \omega \rangle.
\end{equation}

Figure~\ref{fig:cw_laser_effect} shows the eigenenergies and the expectation values of the squared angular momentum for the states $\tilde{J}=0, 1, 2$ of a linear molecule in a cw laser field. All the eigenstates are high-field seeking, as the polarizability interaction is purely attractive. The field enhances most readily the $\langle \mathbf{J}^2 \rangle$ value of the $\vert 0,0 \rangle$ and $\vert 1,1 \rangle$ states and reduces the $\langle \mathbf{J}^2 \rangle$ value of the $\vert 2,0 \rangle$ and $\vert 3,1 \rangle$ states. This is connected with the ordering of the eigenstates, governed by the formation of tunneling doublets and the alignment of the induced dipole, which causes the energy of the eigenstates with a given $\tilde{J}$ to be generally higher for $|M|= \tilde{J}$ than for $M= 0$.

\section{Results for weakly bound $^{87}\text{Rb}_2$}

Here we evaluate the energy shifts of the rotational levels hosted by the highest vibrational states of the $^{87}\text{Rb}_2$ molecule. Table~\ref{table:Param} lists the relevant parameters of the $X^1\Sigma$ and $~a^3\Sigma$ electronic states of the dimer.

The energy shifts for the highest vibrational states of  $^{87}\text{Rb}_2$ in its lowest singlet and triplet electronic states are listed in Tables~\ref{table:cw},~\ref{table:Short}, and~\ref{table:VeryShort} for a cw and a pulse laser field. Note that the intensity of the pulsed laser field is chosen to exceed the cw laser intensity by a factor of ten. However, our estimate of the Keldysh parameter~\cite{KeldyshJETP65} suggests that no appreciable field ionization of Rb$_2$ will take place at laser intensities
up to $10^{12}$ W/cm$^2$.

One can see that for a cw laser field, the shift $\Delta E_r$ of the choice rotational levels exceeds by far their vibrational shift $\Delta E_v$. This is due to the $\omega_\perp$ term of the reduced Hamiltonian~(\ref{ham1}), which is always greater than the $\Delta \omega$ term, responsible for the vibrational shift. Therefore, for the rotational states considered, a cw laser field shifts only slightly the host vibrational level upward, but pushes significantly all the rotational levels downward. It is worth noting that for the weakest-bound species even a small $\Delta E_v$ value may be enough to push the vibrational state out of the effective potential  and thus dissociate the molecule. The requisite critical values, $\langle \mathbf{J}^{*2} \rangle$, of the angular momentum needed for dissociation, cf. ref. \cite{LemFriPRArapid09}, are listed in Table~\ref{table:Param}.  
The twofold effect brought about by long laser pulses which interact with the molecule adiabatically is the same as that of a cw laser field. 

The twofold effect of a pulsed laser field is summarized in Tables~\ref{table:Short} and~\ref{table:VeryShort}. Since the $g(t) \omega_\perp$ term is nonzero only when the field is on, it does not affect the rotational energy $\Delta E_r$ imparted nonadiabatically. For $J=0$ and $1$, both the $\Delta E_r$ and $\Delta E_v$ values are always positive, resulting in a positive total energy shift $\Delta E_t$. For weakly bound states, the vibrational shift alone may be enough to dissociate the molecule, as exemplified by the $X^1\Sigma,~v=124, \vert 0, 0 \rangle$ and $a^3\Sigma,~v=40, \vert 0, 0 \rangle$ states subjected to a 15~ns pulse and marked as $>E_b$ in Tables~\ref{table:Short} and~\ref{table:VeryShort}. The $\Delta E_r$ and $\Delta E_v$ can also reinforce each other and push both manifolds upwards, as exemplified by the $J=1$ states in both the $X$ and $a$ potentials.

\section{Conclusions and Prospects}
\label{sec:conclusions}

In this paper, we tackled the problem of how a nonresonant laser field affects the energy levels of a linear molecule in different temporal and intensity regimes. We were able to evaluate the twofold effect that a field of intermediate intensity (up to $10^{12}$ W/cm$^2$) exerts on the molecule, namely a shift of the molecule's vibrational and rotational manifolds. Both types of shift were found to be either positive or negative, depending on the initial rotational state acted upon by the field. 

A cw laser field or a laser pulse long compared with the molecule's rotational period (adiabatic interaction) imparts a rotational energy shift which is negative and exceeds the concomitant positive vibrational shift  by a few orders of magnitude. The rovibrational states are thus pushed downward in such a field. However, for the weakest-bound vibrational states, even a slight positive vibrational shift induced by the nonresonant field may suffice to cause predissociation.

In the case of a nonresonant pulsed laser field that interacts nonadiabatically with the molecule (pulse duration less than the rotational period), the rotational and vibrational shifts are found to be of the same order of magnitude. Since both shifts are solely due to the imparted angular momentum $\langle \mathbf{J}^2 \rangle$, they always have the same sign, e.g. positive for the $\vert 0, 0 \rangle$ initial state and negative for the $\vert 2, 0 \rangle$ initial state. The nonadiabatic energy transfer occurs most readily at an intermediate pulse duration, $\tau \sim 0.1~\tau_r$, and vanishes in the highly nonadiabatic regime attained when  $\tau \sim 0.01~\tau_r$. We applied our treatment to the much studied $^{87}$Rb$_2$ molecule in the last bound vibrational levels of its lowest singlet and triplet electronic states.  Our calculations indicate that $15$~ns and $1.5$~ns laser pulses of an intensity in excess of $5\times10^9$~W/cm$^2$ are capable of dissociating the molecule due to the vibrational shift (in the absence of additional vibrational-rotational coupling). Lesser shifts can be used to fine tune the rovibrational levels and thereby affect collisional resonances by the nonresonant light.

The twofold effect may be discernible spectroscopically, especially for  transitions involving states with opposite shifts, such as the $\tilde{J}=1 \to \tilde{J'}=2$ transition, cf. Fig.~\ref{fig:DeltaE}. The most pronounced spectral shift, of about 10 MHz at $I \approx 10^{9}$ W/cm$^2$ , is expected to occur for the two-photon $\tilde{J}=0 \to \tilde{J'}=2$ transition. These nonresonant field effects complement those tackled previously, e.g., in ref.~\cite{FriHerCPL96}, and could be discerned by post-pulse quantum beat spectroscopy \cite{CarterHuberQBS00}.

\section{acknowledgements}
We thank Christiane Koch and Ruzin Agano\u{g}lu for discussions and Gerard Meijer for encouragement and support.

\bibliography{References_library}
\end{document}